\documentclass[preprint,amsmath,3p,twocolumn]{elsarticle} 

\usepackage{hyperref}   
\usepackage{amsmath}    
\usepackage{graphicx}                    
\usepackage{epsfig}
\usepackage{xcolor}

\newcommand{\be}{\begin{equation}}
\newcommand{\ee}{\end{equation}}

\begin{document}

\title{Probability distribution function of crossover frequency of operational amplifiers 
}

\author[1]{Todor~M.~Mishonov\corref{cor1}}
\ead{mishonov@bgphysics.eu}

\author[2]{Vladimir~G.~Marinov}
\author[2]{Victor~I.~Danchev}
\author[1]{Emil~G.~Petkov}
\author[2]{Aleksander~P.~Petkov}
\author[1,3]{Iglika~M.~Dimitrova}
\author[2]{Vassil~N.~Gourev}
\author[1]{Nikola~S.~Serafimov}
\author[4,5]{Aleksander~A.~Stefanov}

\author[1]{Albert~M.~Varonov\corref{cor2}}
\ead{varonov@issp.bas.bg}

\cortext[cor1,cor2]{Corresponding author}

\address[1]{Georgi Nadjakov Institute of Solid State Physics, Bulgarian Academy of Sciences,\\
72 Tzarigradsko Chaussee Blvd., BG-1784 Sofia, Bulgaria}
\address[2]{St.~Clement of Ohrid University at Sofia,
5 James Bourchier Blvd., BG-1164 Sofia, Bulgaria}
\address[3]{Faculty of Chemical Technologies, University of Chemical Technology and Metallurgy, \\
8 Kliment Ohridski Blvd., BG-1756 Sofia}
\address[4]{Faculty of Mathematics, St.~Clement of Ohrid University at Sofia,
5 James Bourchier Blvd., BG-1164 Sofia, Bulgaria}
\address[5]{Institute of Mathematics and Informatics, Bulgarian Academy of Sciences, \\
Acad.~Georgi Bonchev Str., Block 8, BG-1113 Sofia, Bulgaria}

\date{1 December 2020, 18:22}

\begin{abstract}
For the first time the cumulative distribution function of the crossover frequency of 
a contemporary operational amplifier ADA4898-2 is experimentally studied.
Using a USB Lock-In amplifier, which allows automatic frequency sweep of the current response of a non-inverting amplifier with significant static amplification, we measure the crossover frequency of 200 samples of ADA4898-2 operational amplifiers.
This new method gives a significant advantage in accuracy and speed of study of every operational amplifier.
The theory we use is based on the universal relation between time dependent output and input voltages.
This common relation for all operational amplifiers is applicable for frequencies much smaller than the 
crossover frequency and the frequencies of non-dominant poles.
In other words, this approximation is adequate, when an operational amplifier is included in a circuit with significant amplification.
\end{abstract}

\maketitle

\section{Introduction}
\textcolor{black}{
The crossover frequency of the operational amplifiers $f_0$
is one of the most important parameters given practically in all manufacturers specifications.
Often for important parameters manufacturers give histograms
of their distribution demonstrating the reliability of the operational amplifier 
in technical applications.
However, $f_0$ is an exception and
according to the best we know,
statistical properties have never been presented;
perhaps because the known methods require significant time for
precise measurements.
The purpose of the present research is to represent a fast method
for measurement of $f_0$ which requires use of a contemporary 
lock-in voltmeter with automatic frequency sweep.
This method allows to study a big series of contemporary
low noise and high-frequency operational amplifiers.
The exact knowledge of $f_0$ is necessary for calculation
of passband width of amplifiers 
used for precise measurement 
of small noises for study of Jonson-Nyquist thermal noise\cite{k_b}  
and Schottky shot noise \cite{q_e}.
Another technical application is, for example, 
a resonator made by general impedance converter~\cite{GIC}.
In this device the resonant frequency explicitly depends on $f_0$.
An extended list of articles related to the applications and use of $f_0$ we have given in the cited references \cite{k_b,q_e,GIC}.
In the next section we recall all mathematics necessary
to explain our method for determination of $f_0$.
In order to explain in which plot the experimental data processing 
can be reduced to reliable linear regression,
it is necessary to analyze basic formulae describing the work of operational amplifiers.
}

\section{Recalling some basic properties of operational amplifiers}

The non-inverting amplifier is one of the most frequently used circuits in electronics.
We will not cover it with thousands of references but we will confine our attention to a singular example. The specifications of ADA4817 \cite[Eq.~(4)]{ADA4817} gives the frequency-dependent amplification $\Upsilon(\omega)$, which is the ratio between the output $U_0$ and input $U_{\mathrm{I}}$ voltages, as 
\be
\label{NIA}
\Upsilon(\omega)=\frac{U_0}{U_{\mathrm{I}}}=\frac{1}{\left(\dfrac{r}{R+r} \right)+s \tau_{_0}}, 
\ee
where (after some change in notation): $r$ is the gain resistance; $R$ is the feedback resistance; $\omega$ is the angular frequency and $\tau_{_0}$ is the time-constant, parameterized by the crossover frequency 
\be
f_0=\frac{1}{2 \pi \tau_{_0}}, \qquad f = \frac{\omega}{2 \pi},  \qquad s=\mathrm{j} \omega,
\ee
and $\mathrm{j}$ is the imaginary unit. 
In the physics and techniques of waves a plane wave has an amplitude 
$\propto \mathrm{e}^{\mathrm{j}(\omega t-\mathbf{k}\cdot\mathbf{r})}$.
Actually in Ref.~\cite[Eq.~(4)]{ADA4817} Eq.~(\ref{NIA}) is written as (this well-known result does not depend on loading
until operational amplifier is in linear regime)
\be
\label{Adelina}
\frac{U_0}{U_{\mathrm{I}}}=
\frac{2\pi\times f_\mathrm{CROSSOVER}(R_G+R_F)}
{(R_G+R_F)S+2\pi\times f_\mathrm{CROSSOVER}R_G},
\ee
for $f_0/G_0\ll f\ll f_0$
with apparent correspondence to the notations:
undefined $S=s$, $f_\mathrm{CROSSOVER}=f_0$,
$R_F=R$, and $R_G=r$.
In other words, very often $f_\mathrm{CROSSOVER}$ is not exactly -3dB open loop frequency
but the parameter describing the low frequency response $f_0$.
The derivation of these well-known formulae is elementary 
and we have reproduced it in the appendices of Refs.~\cite{k_b,NDC}.
It is interesting to compare this result and the famous amplification formula from the well known monograph~\cite[chap.~3, Fig.~3.6b, Eq.~(3.12)]{Albert:87}
\be
\label{Albert}
\Upsilon(\omega)=\frac{U_0}{U_{\mathrm{I}}}=\frac{1}{\left(\dfrac{r}{R+r} \right)+G^{-1}(\omega)}, 
\ee
where $G(\omega)$ is the frequency-dependent open-loop gain of the operational amplifier, 
used in the relation 
\be
\label{GlavnoUravnenie}
U_+ - U_- = G^{-1}(\omega) U_0
\ee
between its input and output voltages.
The function $\Upsilon(\omega)$ is a complex function from a real argument.
The comparison between Eq.~(1) and Eq.~(3) shows that 
\be
\label{low_frequency_gain}
G^{-1}(\omega)=\mathrm{j}\omega \tau_{_0}+\frac{1}{G_0}
=s\tau_{_0}+\frac{1}{G_0}, \quad s=\mathrm{j}\omega,
\ee
where the DC open-loop gain $G_0 \gg 1$ and the second term is negligible in any high-frequency applications.
Most of the modern operational amplifiers, and especially the general-purpose and fast setting types,
have a gain frequency response $G(\omega)$ which can be accurately approximated by the simple one pole response given by the formula above. 
This equation can be found in numerous monographs, for instance cf. Ref.~\cite[Chap.~2, Eqs.~(2.7a) and (2.7b)]{Dostal} and Eq. (2.7a) therein.
The approximation
\be
\label{approximation}
G^{-1}(\omega)\approx\mathrm{j}\omega \tau_{_0}=s\tau_{_0}
\ee
substituted in Eq.~(\ref{Albert}) gives the well-known formulas Eq.~(\ref{NIA}) and
Eq.~(\ref{Adelina}).

Here we suppose that all voltages are sinusoidal $U \propto \mathrm{e}^{\,\mathrm{j} \omega t}$.
In time representation for arbitrary time dependence of the voltages $U(t)$
one can introduce notation for the time differentiation $\hat{s}=\frac{\mathrm{d}}{\mathrm{d}t}$
and in this case $s$ is its eigenvalue
\be
\hat{s}\mathrm{e}^{\, \mathrm{j} \omega t} = s\mathrm{e}^{\, \mathrm{j} \omega t}.
\ee
We where unable to find in the literature the time representation of Eq.~(\ref{low_frequency_gain}).

After introducing the notations we can recall the derivation
of the closed loop amplification 
Eq.~\ref{Adelina} and Eq.~\ref{Albert}
of an non-inverting amplifier
depicted after voltage divider in Fig.~\ref{meas}:
\begin{align}&
U_+=U_\mathrm{I},
\quad
U_-=\frac{r}{R+r}\,U_0,
\quad
U_0=(U_+-U_-)\,G,
\nonumber \\&
G=f_0/s,
\quad
s=\mathrm{j}\omega.
\end{align}
We mark the main details:
the input voltage $U_\mathrm{I}$ is applied 
to the (+) voltage input of the operational amplifier,
$U_-$ is the voltage at the (-) input of the operational amplifier,
$U_-$ can be expressed by output voltage $U_0$
by a voltage divider,
output voltage $U_0$ is represented by the 
voltage difference at the input electrodes $U_+-U_-$,
the open loop gain $G$ is frequency $f$ dependent,
and the frequency $s$ is purely imaginary variable;
the algebraic derivation is elementary and given it the textbooks.

\section{Master equation for operational amplifiers}

Practically all textbook in electronics are written in frequency representation,
see for example the nice monograph by Dostal~\cite[]{Dostal}
and some of them have even time domain considerations, see the Laplace transforms considered in Ref.~\cite{Oppenheim}.

The conversion between frequency and time representation is well-known, 
but in some sense the time representation is in the beginning as all fundamental laws are. 
The fundamental laws in physics are related with time evolution.
Imagine a textbook in electrodynamics written in frequency representation.
It is the same in electronics and in many problems we need to use time representation and to solve 
differential equations.
For this purposes we need to make the transition
\be
G^{-1}(\omega)=\frac{1}{G_0}+\mathrm{j} \omega \tau_{_0} \quad \rightarrow \quad \hat{G}^{-1}=\frac{1}{G_0}+\tau_{_0} \frac{\mathrm{d}}{\mathrm{d}t}
\ee
and in such a way we obtain the master equation of all operational amplifiers
\begin{align}
U_+(t) - U_-(t) & =\hat{G}^{-1} U_0(t) \nonumber \\
&=\left ( \frac{1}{G_0}+\tau_{_0} \frac{\mathrm{d}}{\mathrm{d}t} +
\dots\right ) U_0(t).
\label{master}
\end{align}
This equation is well-known to the majority of electrical engineers,
see for example the book by Siebert~\cite{Siebert}
and in this sense we can see
that the purpose of the present work is to determine
this important for all operational amplifiers parameter 
$\tau_0=1/2\pi f_0$.
Here dots means the next terms in the Taylor expansion with respect of time derivatives, which are as a rule negligible when amplification of the circuit is significant.

Despite the fact that this master equation is implicitly present in tens of thousands of publications in electronics, it has never been published in explicit form. 
Perhaps the first hints of such a dependence have been given by Hendrik Bode~\cite{Bode:40}
see also the work by Peterson~\cite{Peterson:34}, but the term operational amplifier had not yet been canonized at that time.
Coining the term operational amplifier~\cite[page~H.16]{Jung:02}, John Ragazzini~\cite[Eq.~(6), Eq.~(7), Eq.~(32)]{Ragazzini:47} summarizes the experience of significant engineering developments from the Manhattan project~\cite{nytimes:88} and therefore it would be deservedly to say that Eq.~(\ref{master}) is ``Manhattan equation'' for operational amplifiers in spite
that it is only one seminal paper. 
Alternative ideas are to say ``master equation'' in sense ruling equation (nothing related to stochastic theory) or ``main equation'' for operational amplifiers.

\section{Applicability of the low frequency approximation}

In more that 51\% of its applications, operational amplifiers are used in closed loop circuit 
with significant amplification much bigger than one. 
In this case the frequency domain is significantly reduced by the amplification.
Only in this well-known approximations are derived the well-known formulas
for frequency dependent amplification of inverting and non-inverting amplifiers
given in the specification of ADA4817 \cite[]{ADA4817}.
In order to clarify the terminology, we have to point out that the low-frequency approximation 
is parameterized by time constant $\tau_{_0}$ or the corresponding frequency
$f_0\equiv1/2\pi\tau_{_0}.$
Confer the relation Open-Loop Gain versus Frequency graphically 
represented in Ref.~\cite[]{ADA4817}.
Only within the framework of this extrapolation we can call $f_0$ as approximately
-3dB bandwidth frequency of crossover frequency.
We follow the language of Ref.~\cite{ADA4817}
but strictly speaking in the present paper we present a method for determination
of the frequency parameter $f_0$ and its Probability Distribution Function (PDF)
for a series of hundred samples. 
PDF is actually the integrated histogram which is often more informative for descriptive statistics.

In the general case, however, the low frequency approximation in frequency representation of
Eq.~(\ref{master}) has to be substituted by an adequate Pad\'e approximant
\be
\label{low_frequency}
G^{-1}(s)\equiv G^{-1}(\mathrm{j} \omega)=\left(\frac{1}{G_0}+s\tau_{_0}\right)\frac{P_n(s)}{P_m(s)},
\ee
where the polynomials in the numerator and the denominator start from a constant term one
$P_n(0)=P_m(0)=1.$
This correction giving the exact formula is negligible, for example,
for a static amplification according to Eq.~(\ref{NIA}) with $R/r>10$.
But of course in the general case the single pole approximation Eq.~(\ref{low_frequency})
is inadequate for high-frequency applications.
In order to have adequate description, we have to know a good parameterized function
$G^{-1}(s)$, which is specific for each operational amplifier.
However, the low frequency approximation used for example in Ref.~(\cite{ADA4817})
is universal.
In the next section we describe a method for determination of $f_0$
and further we represent the statistical distribution of this important for the operational amplifiers parameter.

\section{Experiment with closed-loop circuit for determination of $f_0$}

Perhaps the simplest idea for determination of $f_0$ is 
to use study frequency dependent voltages of an inverting amplifier. 
Substituting $U_-=0$ in Eq.~(\ref{GlavnoUravnenie}) and the approximation
for the frequency dependent open loop gain Eq.~(\ref{approximation})
we easily obtain the approximate formula from the text book by Dostal
~\cite[]{Dostal}
\be
\label{simplest_main}
f_0\approx f\frac{|U_0|}{|U_-|},
\quad\mbox{for}\quad f_0/G_0\ll f \ll f_0.
\ee
However this intermediate approximation cannot be used at very low frequencies where measurements 
have better accuracy.

There is no doubt that the engineers from those times were familiar with operational calculus and complex analysis.
But that is not the case for the contemporary textbooks in electronics addressed to users abhorring complex numbers and differentiation.
This probably is the main reason a method for the crossover frequency $f_0$ measurement not to be described.
A method for determination of $f_0$ described in 
Ref.~\cite{Dostal}
requires significant time for the measurement.
However, despite this textbook explanation as a rule most of the users are unable to determine 
$f_0$, and many manufacturers of operational amplifiers give in their specifications 
values for $f_0$ significantly differing from the measured values.

One of the goals of the current work is a simple method for the determination of $\tau_{_0}$ and $f_0$.
Let introduce the modulus of the complex amplification function
$Y(\omega)\equiv\left| \Upsilon(\omega)\right|$.

Concerning the variable $S$ in the specifications of operational amplifiers 
without an explanation, 
in the engineering zone of Analog devices it is explained as s-domain 
in sense of Laplace  transformation which is an easier and more compact form to work with.
However, from Laplace imaginary frequency we have to make analytical continuation
to real frequencies $\omega$ and thus $s=\mathrm{j}\omega$ is purely imaginary.
In such a way  
when $s$ is purely imaginary, from Eq.~(\ref{NIA}) we obtain
\be
\frac1{Y^2(f)}=\frac{\left|U_\mathrm{I}\right|^2}{\left|U_0\right|^2}
=\dfrac1{\left(\dfrac{R}{r}+1\right)^2}+\frac{f^2}{f_0^2},
\quad \mbox{for}\quad f\ll f_0.
\label{eq:Uf}
\ee
The advantage of this formula Eq.~(\ref{eq:Uf}) for determination of $f_0$ with respect to simplest idea
Eq.~(\ref{simplest_main}) is that we can use exact linear regression up to zero frequency.
In such a way the increased accuracy allows for first time to study such a subtle 
peculiarities as probability distribution function of crossover frequency for a big series of contemporary operational amplifiers.

For a small enough frequency the amplification is almost equal to the static value ($f=0$)
$Y_0=R/r+1.$ 
Then we measure the frequency  $f_{1/n}$, 
for which the amplification $Y(f_{1/n})$ is $n$ times smaller 
than $Y_0$, 
i.e. $Y(f_{1/n})=Y_0/n$.
In this way the crossover frequency can be determined as 
\be
f_0=\frac1{\sqrt{n^2-1}} \left( \frac{R}{r}+1\right) f_{1/n}.
\label{f_1/n}
\ee

The first experimental data for $n=2$ in the $Y^{-2}$ versus $f^2$ plane for ADA4898-2~\cite{ADA4898} operational amplifier is presented in Ref.~\cite{AIP}, where using a dozen ADA4898-2 we obtained $f_0=(46.6\pm1.3)$~MHz.

Most frequently used notion in this direction is the 
closed loop $f_\mathrm{-3dB}$ frequency defined by the 
equation
\be
Y^2(f_\mathrm{-3dB})=\frac12\approx 10^{-3/10}.
\ee
The substitution $n^2=2$ in Eq.~(\ref{f_1/n})
gives
\be
f_\mathrm{-3dB}=\frac{r}{R+r}f_0=\frac{f_0}{Y_0}
\ee
in agreement with Eq.~(9) of Ref.~\cite{ADA4817},
which reveals that crossover frequency $f_0$
used in this data sheet 
has to be interpreted as extrapolation parameter 
describing low frequency behavior of circuits with
operational amplifiers.

\begin{figure}[h]
\centering
\includegraphics[scale=0.2]{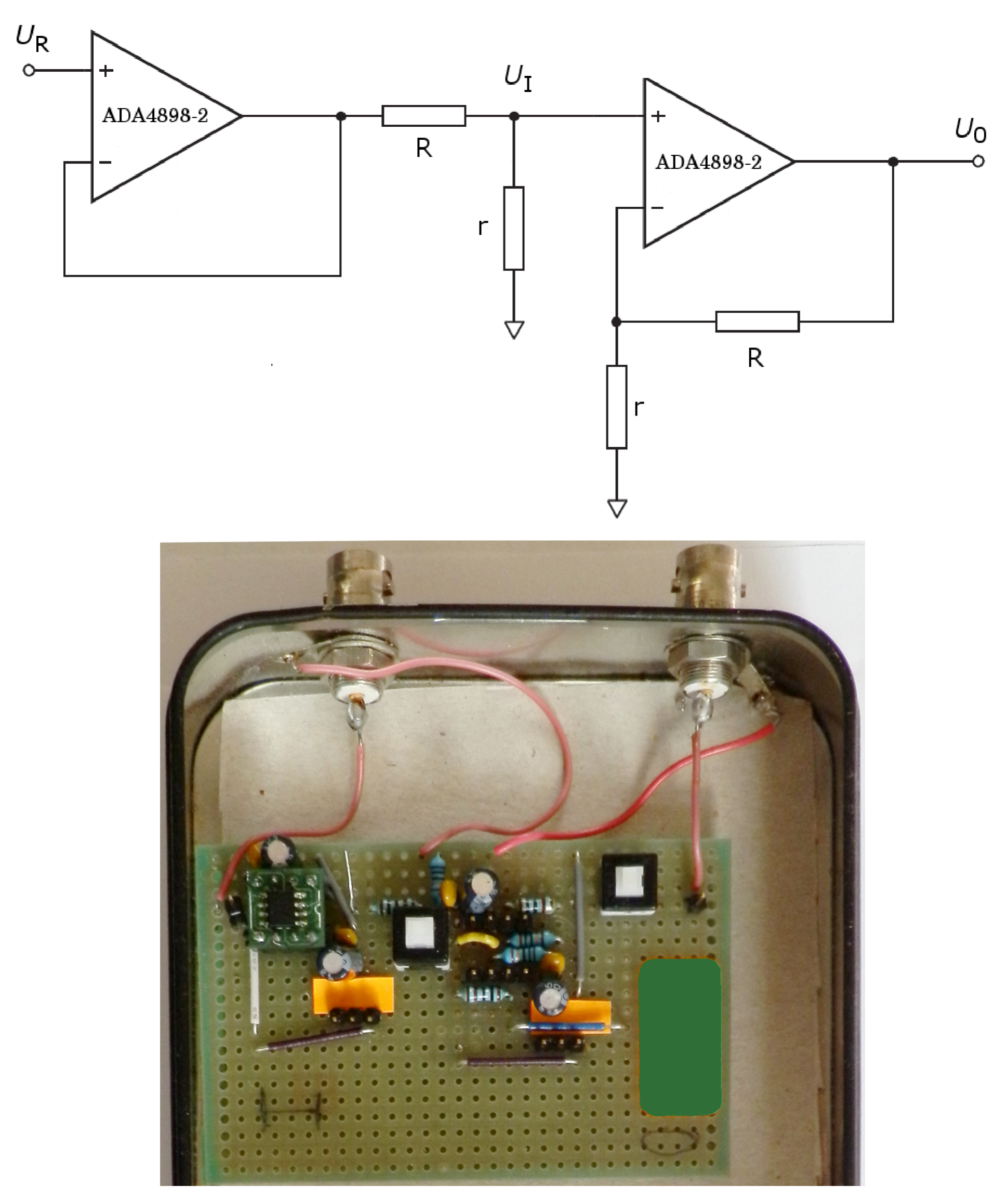}
\caption{Upper:
A schematic of the circuit for the crossover frequency $f_0$ measurement with an Anfatec Lock--in amplifier.
The circuit consists of a repeater with a ADA4898-2 operational amplifier, followed by a voltage divider with $R_1=R=1 \; \mathrm{k}\Omega$ and $R_2=r=10 \; \Omega$ and a non-inverting amplifier with the ADA4898-2 whose $f_0$ is to be measured.
The Lock--in reference signal is $U_\mathrm{I}$ and the output signal going back to the Lock--in is $U_0$.
The power supply voltage is omitted for brevity, the used supply voltage for both operational amplifiers is $\pm 9$~V with the described capacitors in the ADA4898 specifications~\cite{ADA4898}.
The static amplification is significant $R/r=100\gg1$ and this ensures applicability of the 
universal master equation for the operational amplifiers Eq.~(\ref{master}). 
\textcolor{black}{Lower: A photograph of the set-up of schematics above 
together with BNC connectors and screening metal box.}
}
\label{meas}
\end{figure}

\begin{figure}[h]
\centering
\includegraphics[scale=.6]{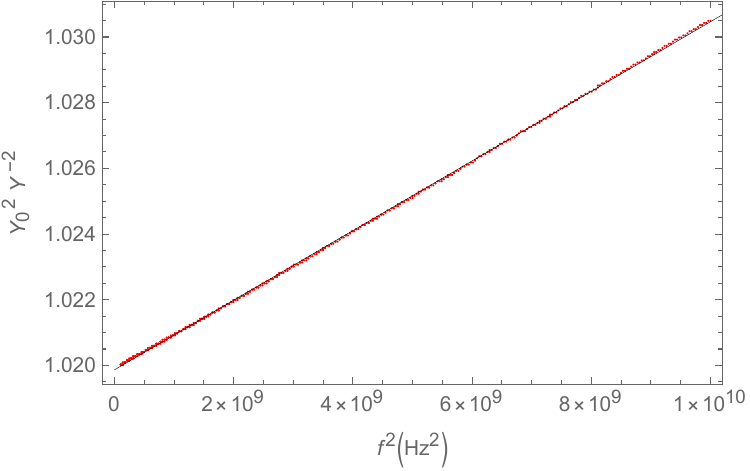}
\caption{A linear regression (solid line) according to Eq.~(\ref{eq:Uf}) from the measurement of one of the 400 ADA4898-2 measured operational amplifiers with an Anfatec Lock--In amplifier and the experimental set-up shown in Fig.~\ref{meas}.
The measurement was made in the frequency range 10-100~kHz with 512 points and the high correlation coefficient $\rho>0.9999$ shows the reliability of the method.}
\label{linear}
\end{figure}

Another set of measurements using the same master equation but now with an Anfatec USB~Lock--in~250~kHz~amplifier~\cite{LockIn} were made.
The Lock--in amplifier applies its reference signal first to a simple repeater circuit with an ADA4898-2 amplifier, then to 100 times voltage divider and finally to a non-inverting amplifier with static amplification 101, which contains the measured ADA4898-2 as is shown in Fig.~\ref{meas}.
The repeater and non-inverting amplifier are placed in a screening metal box from STR8 after shave lotion with both input and output coax cables going to the lock-in amplifier.
Also, the set-up includes an appropriate switch for the selection of only one output signal from the double operational amplifiers at the same time.
Clarifying the measurement itself,
the USB~Lock--in~250~kHz~amplifier~\cite{LockIn} features an option called frequency sweep that allows consecutive measurements within a given range of frequencies.
There are various options for these measurements that are given in the manual~\cite{LockIn} and giving more details here would make this paper look more like an operation manual for a lock--in amplifier.
But in order to alleviate the reproduction of the results presented here, we add a short explanation:
the lock-in applies its reference signal with a given frequency and after the measurement, it automatically changes to the next frequency for the next measurement.
This frequency sweep allows quickly to obtain the data shown in  in Fig.~\ref{linear}.
\begin{figure}[h]
\centering
\includegraphics[scale=.8]{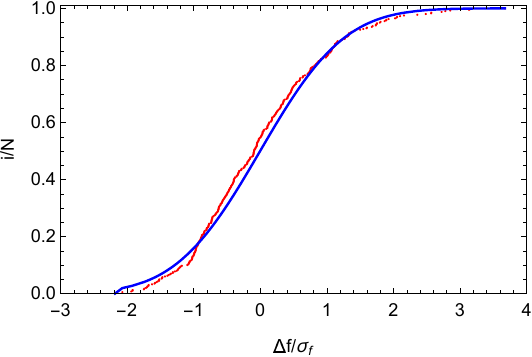}
\caption{For the first time in electronics: probability distribution function (or cumulative distribution) of the measurements of the crossover frequency $f_0$ for 200 ADA4898-2 (double) amplifiers.
The experimental data are shown with the dots and fitted error function is shown by a continuous line.
The abscissa is the dimensionless deviation from the mean value $\Delta f/\sigma_f$, where $\Delta f=f_\mathrm{i}-\overline{f}$, $\overline{f}=97.73$~MHz is the average frequency and $\sigma_f=\langle (f_\mathrm{i}-\overline{f})^2 \rangle=1.62$~MHz is the standard deviation (standard error), where the angle brackets mean division of the sum in the brackets by $N-1$, where $N=400$ is the number of the measured samples.
The ordinate is the ratio between the index of the ordered measurement $\mathrm{i}$ and the total number of measurements $N$.
The difference between the experimental distribution (dots) and analytic fit (line), i.e. the Kolmogorov criterion~\cite{JohnLeon,Bickel} reveals that the distribution is normal within acceptable accuracy,
a linear regression of both distributions has a correlation coefficient 0.996.
Finally to conclude $\sigma_f/\overline{f}=1.66\%$.}
\label{kolmogorov}
\end{figure}
The high correlation coefficient $1-\rho\ll 10^{-3}$ 
of the typical experimental data for the frequency sweep 
depicted in Fig.~\ref{linear} reveals the excellent accuracy of Eq.~(\ref{eq:Uf})
for the studied frequency domain.
Eq.~(\ref{eq:Uf}) is based on the well-known frequency response of the nonlinear amplifier Eq.~(\ref{Adelina}), which is actually 
a consequence of the time dependent equation Eq.~(\ref{master}).
The slope of this linear dependence gives the searched parameter $f_0$
describing low-frequency behavior of operational amplifiers.
Using the automatic sweep of the USB lock-in is easily to measure hundreds such dependence.
Having this set-up and performing $N=400$ measurements (200 ADA4898-2), 
the obtained results for $f_0$ are systematized in the cumulative distribution function
(or probability distribution function according other book on statistics)
is represented in Fig.~\ref{kolmogorov}.

The average cutoff frequency $\overline{f_0}=97.73$~MHz and its standard deviation $\sigma_f=\langle (f_\mathrm{i}-\overline{f})^2 \rangle=1.62$~MHz and $\sigma_f/\overline{f}=1.66\%$.
Finally to add some words for the actual measurement with the Lock--In amplifier.
The Anfatec USB~Lock--In~250 features a sweep frequency, which allows us to change the frequency of the reference signal and to measure its amplification by the non-inverting amplifier as a function of the frequency.
In this way, we actually perform the same measurements as in Ref.~\cite{AIP} (the ones with the generator and the oscilloscope) but with 512 points divided in a frequency range 10-100~kHz.
For each single amplifier we perform the described in Ref.~\cite{AIP} linear regression with 512 points and obtain its crossover frequency $f_\mathrm{i}$ as is shown in Fig.~\ref{linear}.
The lowest correlation coefficient of all $N=400$ linear regressions is 0.99935, which is an order of magnitude lower than all the rest larger than 0.9999.
Let us give some details of the experimental data processing.
After every frequency sweep of the USB lock-in
we obtain the linear dependence 
represented in Fig.~\ref{linear}.
The linear regression gives the slope which determines the crossover frequency $f_0$
for every operational amplifier.
If necessary, the measurement can be repeated.
After that $f_0$ for all operational amplifiers are ordered by its value 
$f_0^{(1)}\le f_0^{(2)}\le f_0^{(3)} \dots \le f_0^{(i)} \dots \le f_0^{(N)}$
and in such a way we obtain the abscissas of Fig.~\ref{kolmogorov}.
The ordinates are sequential numbers divided by total number which are $\le 1$.
The derivative of the cumulative distribution function represented in Fig.~\ref{kolmogorov}
gives the differential distribution function,
and the Kolmogorov criterion gives that for the studied series of ADA4898-2
the distribution is Gaussian.
We have also to point out  that device variation 
and errors of measurement are irrelevant for the problem we study. 
The variations of the $f_0$ are of order of 1\%, 
but the lock-in measures voltages with 4 digit accuracy
and measurements are reproducible. 
The frequency is fixed even with higher accuracy and in such a way
the determined variation of $f_0$ for the studied operational amplifier ADA4898-2 is
a reproducible final result.

Every approximation has of course its limits of applicability.
The single pole approximation
\be
G(\omega)\approx \frac1{\mathrm{j}\omega \tau_{_0}+\frac{1}{G_0}}
=\frac1{s\tau_{_0}+\frac{1}{G_0}}
\ee
can be used for engineering design only when the circuits are used in the frequencies 
much lower than the frequencies of non-dominant poles and zeros;
roughly speaking when static amplification $R/r\gg1.$
Let us analyze a typical example depicted in Figure 19 of Ref.~\cite{ADA4898}.
One can see a pronounced phase dip around 20~MHz and a similar behavior of the amplitude.
In such a way the  single pole approximation for ADA4898
\be
\label{single_pole}
G(\omega)\approx \frac1{\mathrm{j}\omega \tau_{_0}}
=-\frac{\mathrm{j}}{\omega \tau_{_0}},
\quad \mathrm{j}=\exp\left(\mathrm{j}\frac{\pi}2\right),
\quad \frac{\pi}{2}=90^\circ
\nonumber
\ee
is well applicable for frequencies $f\ll$~20~MHz, say $f<$1~MHz
as are the first experimental data~\cite{AIP}.
The phase in this region is approximately 
$-90^\circ$ Ref.~\cite[Figure 19]{ADA4898}.
On the other side in the specification of ADA4817 
Ref.~\cite[Figure 31]{ADA4817} one can easily see the textbook like behavior
$0^\circ$ phase for $f\ll f_0/G_0$ and $-90^\circ$ for the frequencies $f_0/G_0 \ll f\ll f_0$,
cf. Ref.~\cite[]{Dostal}.

Only for low frequencies $f\ll f_0$ all operational amplifiers can 
be described by the universal Manhattan equation
parenthesized by a frequency parameter $f_0=1/2\pi\tau_{_0}.$ 
A simple method for determination of this parameter 
is the goal of the present study.

In the specification~\cite{ADA4898}, see Tables~1 and~2,
our parameter $f_0$
corresponds to a
non-inverting amplifier 
with open $R_\mathrm{_G}=\infty$ 
and closed $R_\mathrm{_F}=0$
and $Y(0)=1.$
For such an amplifier, according to Eq.~(9) in \cite{ADA4817} $f_0$
is closed-loop ``--3 dB'' frequency.
In the specification~\cite{ADA4817} 
$f_\mathrm{_{CROSSOVER}}$ is defined as frequency where
the operational amplifier open-loop 
$$|G(f_\mathrm{_{CROSSOVER}})|=1,$$ 
or gain equals 0~dB.
This notion corresponds to our $f_0$ parameter only if the dependence in Fig.~54 in \cite{ADA4817}
is a straight line.
However in the Eq.~(4-9) in this specification~\cite{ADA4817} 
the notion $f_\mathrm{_{CROSSOVER}}$ is used in the sense of our 
$f_0=1/2\pi\tau_{_0}$ notation.
As a parameter with dimension frequency 
describing the low frequency behavior of the electronic circuits with 
amplification $Y(\omega=0)\gg1$, which is the main domain of the 
operational amplifiers market.
Engineers in electronics, who design devices working at frequencies much lower than $f_0$,
for a precise analysis need to evaluate the non-linearity corrections.

\section{Consideration of the results and its applications}

The exact knowledge of the crossover frequency $f_0$ is necessary when we need to precisely determine the non-ideal effects of operational amplifiers, for instance see Ref.~\cite{Finetti:19}.
For instance, in cases when there is a need of an exact calculation of the pass bandwidth of amplifiers with active filters, the Manhattan equation is indispensable.
We hope that our work will induce further tutorial descriptions in AnalogDialogue for the determination of the crossover frequency using graphical representation of the real data for operational amplifiers, 
for example 
$f_\mathrm{_{CROSSOVER}}=410$~MHz for ADA4817~\cite{ADA4817,ADA4898}.

In short, the master equation for the operational amplifiers is present implicitly in electronics development even before the notion \textit{operational amplifier} to be coined but according to the best we know~\cite[Eq.~(2.7a)]{Dostal}, \cite[Eq.~(2.1)]{SPICE}, \cite[Eq.~(2.24)]{Ghausi},\cite[Eq.~(3.47)]{Sun}  and \cite[Eq.(2.1)]{Mohan} this equation has never been published as an equation connecting time dependent input and output voltages; that is why Eq.~(\ref{master}) is a new one.

We apply this equation in the analysis of a simple circuit for measuring of the cut-off frequency of operational amplifiers.
A new equation Eq.~(\ref{eq:Uf}) has been derived and used to develop a new method for the measurement of $f_0$.
Using this new method, we observe significant differences in the values for cut-off frequencies given in the specification by the manufacturer.
Our measurements reveal high reliability, the correlation coefficient is larger than 0.999 and this reveals high accuracy for the applicability of the master equation Eq.~(\ref{master}).
This also means that including higher order terms and second poles in Eq.~(\ref{master}) is not justified since their influence would be hardly detectable and important.

The large difference in the results for $f_0$ obtained by both methods (here and in Ref.~\cite{AIP}) can have several explanations.
The most plausible one is that the first measurement was done with oscilloscope probes for connectivity, which have low pass filters built-in and they introduced spurious frequency deflection.
Another explanation is the screening, however this effect should not introduce a difference 2-3 times but at most around 10\% in the kHz region.

The very high correlation coefficients demonstrate that Eq.~(\ref{eq:Uf}) is a perfectly working consequence of the proposed master equation Eq.~(\ref{master}).
Our method reveals that contemporary operational amplifiers have only 1-2\% dispersion of the cut-off frequency. 
We made analogous measurements using many other operational amplifiers and obtained satisfactory work of our method based on Eq.~(\ref{eq:Uf}).

In conclusion, we suppose that our method for determination of the cut-off frequency based on the Manhattan equation for the operational amplifier can become a convenient tool for engineers developing new circuits and measurement devices\cite{Hoja:08,Pop:20}. 
Our set-up can be built in every lab and can give many examples of derivation of new results using
Eq.~(\ref{master}).
If the culture for measurement of cut-off is popularized, the manufacturer will be more precise when giving this parameter in the specifications, for some purposes 20\% difference is significant.

In order to avoid misunderstandings, we have to emphasize that the frequency parameter $f_0$, which determines circuits with high static amplification describes only the universal low frequency dependence of operational amplifiers. 
We present a reliably working method for determination of this important parameter and perhaps for the first time we present statistical properties of $f_0$ for
a contemporary low-noise operational amplifier. 
For circuits with significant static amplification $\gg1$ formulas with $f_0$ give an adequate description of their frequency behavior. 
In comparison with the well-described methods in the textbooks, our method is based on a simple circuit
and automatic frequency sweep gives the possibility for fast linear regression and analyzing statistics of 
$f_0.$ 
Probability distribution function for 
$f_\mathrm{CROSSOVER}$ can become an important part of the specifications of operational amplifiers.

For the first time in electronics we present experimental data for probability distribution function of the crossover frequency of operational amplifiers. 
We use 400 (200 double) ADA4898-2 low noise amplifiers. 
We present an innovative method for determination of crossover frequency which requires an USB Lock--In with automatic frequency sweep.
Our method is based on 
a differential equation relating the voltage at the output of an operational amplifier $U_0$ and the difference between the input voltages ($U_{+}$ and $U_{-}$) which has been derived. 
The crossover frequency $f_0$ is a parameter in this operational amplifier master equation.
The formulas derived as a consequence of this equation find applications in thousands of specifications for electronic devices but as far as we know, the time dependent equation has never been published.
Actually, the master equation of operational amplifiers can be found in the seminal article by
Ragazzini, Randall and Russell~\cite[Eqs.~(6), (7), (32)]{Ragazzini:47},
but for more than 70~years it was not analyzed and cited in journals and specifications of 
operational amplifiers.
During World War II, John Ragazzini was involved in the Manhattan Project~\cite{nytimes:88}
working on significant projects in the field of electronics and therefore it would be deservedly to say that the master equation we propose is ``Manhattan equation'' for operational amplifiers.
The exact knowledge of the crossover frequency $f_0$ is necessary when we need to precisely determine the non-ideal effects of operational amplifiers.
For instance, in cases when there is a need of an exact calculation of the pass bandwidth of amplifiers with active filters, the Manhattan equation is indispensable.

\section{Discussion and Conclusions}
\subsection{Repeating the terminology. Discussion}
Ones again let us clarify the terminology.
First let us state some well definitions~\cite{ArtKalb}.

1)~-3dB bandwidth/frequency $f_\mathrm{-3dB}$: 
This is the frequency where the closed-loop gain of the amplifier system reaches -3dB. Normally this is specified in a non-inverting gain of 1 (unity feedback) configuration. 
The -3dB bandwidth typically results in the highest bandwidth specification 
and is often used for high-speed amplifiers. 
-3dB bandwidth is independent of loading, we suppose high enough loading impedance.

2)~Unity-gain bandwidth/frequency (UGF)  $f_\mathrm{UGB}$, 
is the frequency where the open-loop gain of the amplifier reaches 0~dB (gain of 1). 
Being an open-loop measurement, 
the feedback configuration should not affect the number when done properly. 
Again,  $f_\mathrm{UGB}$ does not dependent on loading.

3)~Gain-bandwidth Product (GBWP)  $B_\mathrm{GBWP}$ 
is the product of the open-loop gain and frequency measured at a specified frequency (normally listed in the conditions). 
This number gives the frequency at which the amplifier's open-loop response would cross unity when extrapolated out. 
This number is often useful for applications 
where an amplifier is used in a higher gain configuration. 
It will be less dependent on loading conditions.

Now that the definitions are out of the way, 
we can discuss again notions and notation of our work.
and to juxtapose some similar definitions.

First, it is generally best to familiarize oneself 
with the open-loop gain 
and phase plots provided in a data-sheet. 
For the ADA4898-2, this is provided in 
Ref.~\cite[Fig.~19]{ADA4898}. 
Looking at the open-loop gain and phase allows a discerning 
eye to see a number of non-idealities. 
These non-idealities will often introduce performance limitations. 
For example, looking at the ADA4898-2, 
one can see the phase dips around 20~MHz 
and then comes back up again. 
A similar behavior with the magnitude is also visible, 
but the phase behavior is generally more pronounced. 
This dip indicates that a zero has been introduced into the compensation to increase the phase margin while maintaining a high unity-gain crossover point. 
This is acceptable in some applications and not in others. 

Contrary to this the modulus of open loop gain
\begin{align}
A(f)\equiv|G(\omega)|\approx\frac{f_0}{f},
\quad\mbox{for}\quad
\frac{f_0}{G_0}\ll f \ll f_0
\end{align}
of ADA4817-2 represented in 
Ref.~\cite[Fig.~54]{ADA4817} demonstrates
textbook like behavior $A= f_0/f$,
and in the logarithmic plot ($\log A$ versus $\log f$)
the experimental data are approximated with a
mathematical straight line.
Within this acceptable approximation for ADA4817 we have
$f_\mathrm{CROSSOVER}\approx f_\mathrm{UGF}$;
for or other operational amplifiers 
$f_\mathrm{CROSSOVER}$ denoted in the present paper
by $f_0$ has to be defined as linear 
extrapolation from low frequency domain.
The non-idealities are irrelevant for this linear approximation.
For example, if we consider a non-inverting amplifier
built with ADA4898\cite{ADA4898}
if $f_\mathrm{-3dB}\ll20$MHz all equations given
in the present paper and Ref.~\cite{ADA4817} are correct.

For some filter applications perhaps the operational 
amplifiers can be used for frequencies comparable to
$f_\mathrm{UGF}$ but in the many applications 
operational amplifiers are used to create amplifiers
with closed loop gain much bigger than one.
In this case in order to project
the desired system applications, we need to
calculate reliably the frequency behavior for 
frequencies much smaller than $f_\mathrm{UGF}$.
Actually for the most of applications of operational amplifiers
working frequencies are much lower than 
$f_\mathrm{UGF}$.
Here we wish to emphasize the all formulas Eqs.~(4-9)
for frequency dependent amplification
for inverting and non-inverting amplifier
given in Ref.~\cite{ADA4817} are reliable
low frequency approximations for frequencies
$f\ll f_\mathrm{UGF}$, and DC amplifications 
$Y_0\approx R_\mathrm{F}/R_\mathrm{G}\gg 1.$

Usually an engineer selecting an amplifier 
consulting with the price 
makes the choice for which the amplifier
somehow approximates the ideal amplifier for the projected
device.
For example analyzing crossover frequency 
$f_\mathrm{CROSSOVER}$
one can select an amplifier that has more than enough bandwidth and the appropriate gain/phase behavior to meet desired system specifications. 
The purpose of the present study however is completely
different, we do not construct some device 
but examine an important for all operational 
amplifiers parameter $f_0$ defined by low frequency 
extrapolation for which all equations Eqs.~(4-9) Ref.\cite{ADA4817} are correct.
If we need to construct some device working 
at frequencies comparable with $f_0$,
one can recommend $\pm$20\% margin on your bandwidth specification, more or less depending on sensitivity to the bandwidth. 

If the device is sensitive to percent-level variations of the component, we have not left enough engineering margin
and for low frequencies we have to determine $f_0$
precisely.
This was the case when for university education purposed we
constructed amplifier for demonstration of 
Johnson-Nyquist~\cite{k_b} and Schottky noise~\cite{q_e}.

With respect to collecting statistics, 
our method is fine for measurement of
$f_\mathrm{CROSSOVER}$ by extrapolation,
which describes the low frequency behavior of the circuits.
If one want statistics on the $f_\mathrm{UGF}$, 
it is indispensable to measure the point where the gain crosses unity; there is no substitute. 

Analogously to measure
the -3dB point for closed loop amplification we need to measure where the gain crosses -3dB. 
However for high enough static amplifications 
$Y_0\gg1$ 
\be
f_\mathrm{-3dB}\approx f_\mathrm{CROSSOVER}/Y_0
\ll f_\mathrm{CROSSOVER}, 
\ee
and the non-idealities of frequency dependence the operational amplifier are irrelevant.

Let also specify once-again the measurement setup in more detail in order to reconstruct precisely what we measure
and to alleviate the repeating of the experiment.

The schematic is depicted in upper Fig.~\ref{meas}.
The reference signal $U_\mathrm{R}$ is applied 
to a repeater.
Then signal is significantly decreased $r/(R+r)$
times by a voltage divider; $R/r=100$.
The reduced signal $U_\mathrm{I}$
is applied as the input voltage of and 
nonlinear amplifier whit static amplification
$Y_0=(R+r)/r$.
In such a way at small frequencies $f\ll f_\mathrm{-3dB}$
the output signal $U_0=U_\mathrm{R}$ is equal to the 
reference signal and the schematic restores 
the reference signal.
The output signal $U_0$ is measured by 
a USB lock-in which has the option for automatic
frequency sweep.
In such a way we measure not only 
$f_\mathrm{-3dB}$
but the whole frequency dependence 
of $1/Y^2$ versus $f^2$
represented if Fig.~\ref{linear}.
The lock-in has 5 digits accuracy and simultaneously 
the linear regression is the best method
for experimental data processing when applicable.
That is why we take the liberty to assert
that our method is more faster and accurate than other ideas
described in the literature.

Perhaps it is not necessary to mention that lock-in has
very big input impedance and in this sense the load impedance of the schematic in Fig.~\ref{meas}
has to be considered as infinite.

As we use $R_\mathrm{F}/R_\mathrm{G}=100$
the $f_\mathrm{-3dB}$ is hundred times smaller
that $f_\mathrm{CROSSOVER}$
and the non-idealities of ADA4898 at 20~MHz are irrelevant
for our extrapolation method for determination of
$f_\mathrm{CROSSOVER}$, which is definitely
different from the $f_\mathrm{UGF}$ for the studied 
operational amplifier.

From theoretical point of view 
the well-known master equation Eq.~(\ref{master}) for operational amplifiers is better explained in Laplace 
s-domain~\cite{Siebert}
which is a much easier form to work with - and it is more compact.  
However, we wish to emphasize that in many applications
the frequency has to be consider as complex.
The master equation Eq.~(\ref{master})
can be applied for example to study instability 
of negative impedance converters (NIC)
in order to solve such a practical problems:
in which polarity of (+) and (-) inputs the NIC
ca be used to generate oscillations in parallel and
sequential resonance circuits~\cite{NDC}.
For such purposes the master equation
is used to calculate eigen-frequencies of the electronic circuits.

For many purposes however,
the single pole equation is not enough for system design.
Also the non-dominant poles and zeros could be of critical importance. 
Specifically, all operational amplifiers have a non-dominant pole that limits the bandwidth. 
This introduces a phase shift before 
the unity-gain crossover frequency. 
However, before developing methods for 
reliable determination of non-dominant 
poles and zeros of operational amplifiers,
it is necessary to start with a dominant one
which is the subject of the present study.
A complete understanding of an operational amplifier 
includes calculation and experimental determination of all
poles and zeros of his frequency response.
From an analytical point of view we need to have a reliable 
Pad\'e approximation of $G(s)$.

For many cases the the non-dominant pole is 
much more important to the performance of filters than the exact crossover frequency
and it deserves to put in the agenda of electronics the
development of exact and fast methods for determination of the non-dominant poles.
These non dominant poles and zeros determine
additionally, the other gain/phase wiggles which can drastically impact circuit performance. Specifically, low frequency zeros in the open-loop transfer function tend to introduce long-tail settling effects (doublets).

In this direction the low frequency extrapolation method
for measurement of the crossover frequency is
only the first step in this direction.

As the Kolmogorov criterion and integral probability 
distribution function are not often given in the specification
of the operational amplifiers,
we will represent our results for crossover frequency
as differential probability distribution (PDF).

\begin{figure}[h]
\centering
\includegraphics[scale=.8]{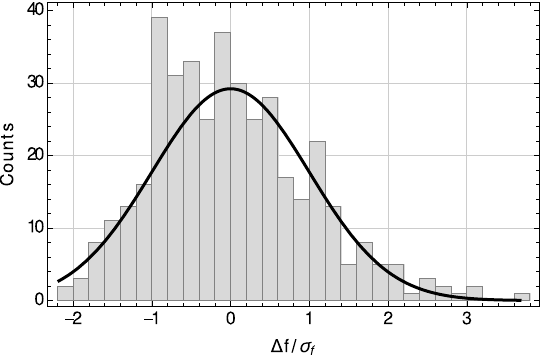}
\caption{The histogram of crossover frequency $f_0$ of the studied ADA4898-2 operational amplifier. 
For $N=400$ samples the number of channels is chosen to be 32. 
The continuous line is the approximating Gaussian.
Confer the histograms of input voltage offset 
and input bias currents of the same operational amplifier
according to Ref.~\cite[Figs.~36, 37]{ADA4898}.
}
\label{histogram}
\end{figure}

In Fig.~\ref{histogram} 
the experimental data from 
Fig.~\ref{kolmogorov} are presented in a histogram. 
As we know for the first time in the literature the statistical properties of the crossover frequency of a operational amplifier are given in a histogram.
The development of a fast and exact method for measuring 
of the crossover frequency was the mission link in this chain up to now and we believe that soon similar histograms can appear in the new editions of the specifications

\subsection{Conclusions}

Our research of $f_0$ is motivated by 
study of fundamental noise and creation of resonance filters for scientific measurements \cite{k_b,q_e,GIC}.
We are going to use the resonance circuit \cite{GIC}
for measurement of Bernoulli effects in superconductors~\cite{m^*_Bernoulli}
and recently predicted by us surface magnetization of superconductors~\cite{m^*_magnetization}. 
Observation of both physical effects leads to the creation of Cooper pair mass
spectroscopy, a new research topic in the physics of the superconductivity.
For those precise applications it is necessary to know the exact value of $f_0$.
That is why represent a reliable method for determination 
of the crossover frequency of operational amplifiers $f_0$.
The knowledge of this parameter is also important for the construction of many 
devices based on operational amplifiers.
A detailed description of the method can encourage performance
of the analogous studies for other operational amplifiers.
And finally we expect the statistical properties of $f_0$ to become 
a routine part in the specifications of the operation amplifiers.
We repeat in short: 
we have developed a new method for determination of the crossover frequency $f_0$
which is orders of magnitude faster and more accurate and can be used by companies
manufacturing operational amplifiers, 
such as Analog Devices and Texas Instruments,
to represent histograms of $f_0$ in the specifications of their products.

\section*{Acknowledgments} 
The authors appreciate the cooperation with Vassil Yordanov in early stages of the present research, 
are grateful to Pancho Cholakov and Ginka Dinekova for suggesting the low noise ADA4898-2 operational amplifier and to Peter~Todorov for providing us the possibility to measure 200 samples.
The authors appreciate the dialog with the Principal Design Engineer/Design Manager Art Kalb through the Analog 
Devices Engineer Zone.

\end{document}